\documentclass{article}
\usepackage{ijcai22}

\usepackage{times}
\usepackage{soul}
\usepackage{url}
\usepackage[hidelinks,hyperindex,breaklinks]{hyperref}
\usepackage[utf8]{inputenc}
\usepackage[small]{caption}
\usepackage{graphicx}
\usepackage{amsmath}
\usepackage{tcolorbox}
\usepackage{amsthm}
\usepackage{booktabs}
\usepackage{algorithm}
\usepackage{algorithmic}
\title{Towards User-Centered Metrics for Trustworthy AI in Immersive Cyberspace}

\author{
Pengyuan Zhou$^1$\thanks{The Corresponding Author: Pengyuan Zhou, E-mail: (pyzhou@ustc.edu.cn)}
\and
Benjamin Finley$^2$\and
Lik-Hang Lee$^{3}$\and
Yong Liao$^1$\and 
Haiyong Xie$^1$\and
Pan Hui$^{2,4}$\and
\affiliations
$^1$University of Science and Technology of China, %\\
$^2$University of Helsinki, %\\
$^3$Korea Advanced Institute of Science and Technology, %\\
$^4$The Hong Kong University of Science and Technology
}

\begin{document}

\maketitle

\begin{abstract}
AI plays a key role in current cyberspace and future immersive ecosystems that pinpoint user experiences. Thus, the trustworthiness of such AI systems is vital as failures in these systems can cause serious user harm. Although there are related works on exploring trustworthy AI (TAI) metrics in the current cyberspace, ecosystems towards user-centered services, such as the metaverse, are much more complicated in terms of system performance and user experience assessment, thus posing challenges for the applicability of existing approaches. Thus, we give an overlook on fairness, privacy and robustness, across the historical path from existing approaches. Eventually, we propose a research agenda towards systematic yet user-centered TAI in immersive ecosystems.
\end{abstract}

\section{Introduction}
\label{sec:intro}

Today, artificial intelligence (AI) methods have shown state of the art performance in many fields and are becoming increasingly widespread in many areas of everyday life, including recommender systems~\cite{cheng2016wide}, health care~\cite{norgeot2019call}, smart factory~\cite{shiue2018real}, financial modeling~\cite{lin2011machine}, marketing~\cite{cui2006machine}, education, science, and commerce~\cite{jordan2015machine}. 
However, such integration also allows AI systems to access large datasets from countless users. AI systems can leverage these datasets along with significant networking and computing power to learn very granular and potentially sensitive user behavior. Additionally, to most users, AI systems appear as black-boxes that provide little insight into their internal decision-making process. This arouses moral concerns surrounding AI systems and especially the trustworthiness of AI for the sake of fairness, privacy, security and system reliability~\cite{adadi2018peeking}. 

To address these concerns and strengthen human trust in AI systems, the area of Trustworthy AI (TAI) has recently seen significant attention from government entities, such as the European Commission~\cite{hleg2019ethics}, United States Department of Defense~\cite{usdod}, China’s Ministry of Science and Technology, and numerous technology giants such as IBM, Google, Facebook.
The major goal of TAI is to ensure the protection of people's fundamental rights while still allowing responsible competitiveness of businesses~\cite{euhumanright}. The term TAI has been around for years and was boosted by the well-known EU guidelines on TAI published in 2019~\cite{hleg2019ethics}. The concept grew rapidly as the number of 
research papers in Google Scholar with the term in the title or abstract increased from 6 to 1,040 over 2017 -- 2021. 

In the TAI domain, TAI metrics are naturally a major issue and critical to accurately measuring the degree of system trustworthiness and the amount of protection offered by AI-enabled technologies. As a quantification of trustworthiness, a TAI metric can be a form of one or multiple system properties or system states to assess trustworthiness.
The appropriate TAI metrics can vary in different domains due to various scenarios, user demand, processed data, adversaries, regulations, and laws. As integration of several related requirements, TAI should contain multiple `dimensions', even though multi-criteria evaluation will lead to increased complexity. 
Despite the large number of case-by-case metrics used in current literature, a comprehensive and systematic outline of TAI focusing on metric selection has yet to be proposed, resulting in challenges for metric choices for non-experts and even professionals. Furthermore, future immersive ecosystems, such as the metaverse, are going to incorporate more complex systems that blend the virtual and physical worlds, and, more importantly, complicated definitions of system performances and user experiences. Therefore, current metrics and metric selection logic might need enhancements and evolution to fit the complexity. 

\paragraph{Related Surveys and Our Scope.}
We acknowledge a number of existing surveys on ethical guidelines for AI~\cite{smuha2019eu}, big data~\cite{mantelero2018ai} and robotics~\cite{torresen2018review}; along with surveys on specific domains (e.g., finance~\cite{lin2011machine}) and types of AI applications (e.g., recommender systems~\cite{gunes2014shilling}). In contrast, this survey explores recent trends and advancements in TAI metric selection across two important domains, and sheds light on the metric selection logic for future user-centered cyberspace. Note that legal compliance (lawful AI) is beyond the scope of this survey. Specifically, we focus on the technical metrics for \textit{fairness}, \textit{privacy}, and, \textit{robustness}. Other TAI requirements, including transparency and accountability, do not have sufficient literature yet for survey purposes and thus are left for future work. Since AI is a socio-technical system instead of a mathematical abstraction, this work requires readers to personalize the metric selection on demand. 

\paragraph{Review Methodology.} The selection of publications was conducted in four steps.
First, we search in Google Scholar for “Trustworthy AI”, and usecase names (e.g., recommendation system) plus TAI requirements (e.g., fairness) and find that relevant publications are spread across multiple scientific journals and conferences. Second, we choose papers primarily from three scientific repositories: ACM Digital Library, IEEE Explore, and Springer.
Third, we try to select the top cited or top-venue (e.g., KDD and WWW) papers if proper papers exist.
Finally, if no proper references are found from the second and third steps, we choose the most feasible references by the searching through Google Scholar with related keywords such as "AI in Networking" and filter by topics with relevant to TAI.

\paragraph{Contributions.} This survey serves as a first effort to outline the critical TAI metrics in current domains, capture the general logic of metric selection, and call for 
advancing user-centered metrics for immersive ecosystems and autonomous metric selections. Specifically, our contributions are threefold.
    First, we describe the importance of the TAI requirements this article focuses on (Section~\ref{sec:requirement}).
    Second, we outline two most fundamental domains and dive into specific use cases therein to summarize TAI metrics in the current cyberspace (Section~\ref{sec:domain}).
    Finally, we summarize the lessons learned from existing TAI metric/ selection methods, and discuss the research agenda for future ecosystems' with TAI as well as the related metric selection methodology (Section~\ref{sec:discussion}).

\section{TAI: User-centered Requirements}\label{sec:requirement}
As mentioned, we focus on \textbf{fairness}, \textbf{privacy}, \textbf{robustness}, and the key metrics to meet these requirements in different domains. In this section, we present an overview of these requirements and list some common metrics. 

\paragraph{Fairness.}
AI, especially machine learning, often presents statistical discrimination due to non-identical data distribution or resources, which leaves certain privileged groups with performance advantages and others with disadvantages. The learning bias, regardless if generated on purpose or accidentally, exacerbates existing resource inequity and further harms fairness in society~\cite{paluck2019contact}. Therefore, the AI community has been putting efforts on measuring the fairness of AI systems to mitigate algorithmic bias.

\paragraph{Privacy.}
Privacy is fundamental yet hard to define explicitly. Nissenbaum~\cite{nissenbaum2004privacy} defines privacy in terms of contextual integrity and contextual information norms dictating how information may be used or shared. As agreed by most researchers, privacy is a multi-dimensional concept~\cite{laufer1973some} and thus is normally assessed via multiple metrics focusing on the exposure of private information. 

\paragraph{Robustness.}
Due to natural variability or dynamic system conditions over time, predicting how future conditions will change might be hard or impossible. The term deep uncertainty, in this scenario, is more proper to describe the issue in AI. Deep uncertainty is defined as a situation in which parties to a decision do not know or cannot agree on how the system (or part of it) functions, the importance of the various outcomes of interest, and/or what the relevant exogenous inputs to the system are and how they might change in the future~\cite{maier2016uncertain}. Under deep uncertainty, typically, multiple environmental state factors, i.e., future conditions, jointly affect the decisions (e.g., policies, designs and plans)~\cite{ben2009robust}, resulting in influences on the considered performance metric (e.g., cost, utility and reliability). Robustness metrics function as a transformation of the performance metrics under these future conditions.

\paragraph{General Rule for Metric Selection.}
In general, the system administrator can follow a series of steps to select the proper metrics: 
    (1) Which requirements of trustworthiness should be assessed?
    (2) Who cares about the issue the most, e.g., the system administrator, the users, the regulators, society etc.?
    (3) Regarding each requirement, who are the major concerned entities for each party (because different parties may have different concerns, e.g., system admins care about performance while users care about privacy), e.g., consistent performance, protected data, equal performance? 
    (4) What is the targeting or common adversary?
    (5) What are the available data resources to compute the selected metrics?
    (6) What is the difficulty and cost of the metric assessment?
    (7) Will the metrics stay valid over time?

Nevertheless, no matter how proper the selected metrics are, they are still only estimations and will not fully encompass all the desired TAI requirements accurately. Additionally, if maximizing the selected metrics became the major model optimization logic, the model may perform well in terms of the TAI metrics but fall short of the original model goal(s). Therefore, periodic user studies are recommended to continuously monitor the system compliance with the TAI metrics. Adaption of the metric selection or standard can be made accordingly to strike a balance between the original goal and trustworthiness.

\section{TAI Metrics in the Existing Cyberspace}
\label{sec:domain}
Although existing TAI guidelines provide assessment metrics, the metrics are often quite high level and thus for individual providers to find or create detailed definitions (the definition of a metric may vary in different fields) is not easy. Moreover, different domains have different performance priorities. Thus, the selection of TAI metrics should naturally also vary. Future ecosystems, with more complexity including in the user experience, only exacerbate these issues and pose more challenges for TAI metric selection. Therefore, we first examine TAI metrics in the current cyberspace and summarize lessons for future ecosystems. In this section, we outline TAI metrics in computing and networking, two basic components of cyberspace that leverage AI.

We find that some metrics are widely used in varying ways across different domains, including regression model metrics, including mean absolute error (MAE), normalized MAE (NMAE), mean squared error (MSE) and root MSE (RMSE), as well as ranking metrics, e.g., hit ratio, precision, recall, specificity (true negative rate), F-score, discounted cumulative gain (DCG), and their varieties. As they usually appear as a group, we refer to them as \emph{regression model metrics} and \emph{ranking metrics} throughout the survey.

\subsection{Computing}
This section focuses on the computing domain. We choose two representative usecases: search engine ranking (SER) and recommendation systems (RecSys), to illustrate the common TAI metrics. These usecases impact huge numbers of users (billions) thus are major concerns in terms of TAI.

Nowadays, SER algorithms consider many factors such as dwell time, page relevance, content quality, and so on. When a search engine presents results, it typically records or calculates such factors based on pre-defined policies, and treat these as implicit proof of user interest. Therefore, the user interaction with the ranking results is critical for training the learning to rank (LR) models. Similarly, RecSys incorporates a number of machine learning techniques and has been widely used in online media platforms, online shopping, and social networks. A RecSys normally collects users' historical choices for supervised learning (e.g., classifying items as recommended or not) or unsupervised learning (e.g., matrix factorization techniques common in collaborative filtering) to learn and predict the user interest in items.

SER and RecSys provide sorted results to users aiming to show results that match the user's search input or recommend items of user interest. However, when a user clicks on the top-ranked link on the result page determining whether the selection is simply because the result is top ranked/recommended or really the most relevant/interesting is difficult, as discovered by the researchers~\cite{joachims2017accurately}. Since top results have higher chances of being selected thus potentially increasing revenues, it is important to guarantee the trustworthiness of the presented results for the benefit of users. 

\paragraph{Fairness}
Fairness is a major concern for SER and RecSys. For instance, SERs sometimes are found to systematically favor certain sites over others in the results, thus distorting the objectiveness of the results and degrading user trust~\cite{tavani2012search}. Additionally, in RecSys, the number of recommendations is often fixed, therefore there are strong incentives to promote products with greater commercial benefit instead of fair recommendations based on ethical data mining.

Besides commercial incentives, implicit biases based on ethnicity, gender, age, community and so on, that widely exist in society, are often reflected in the big data collected from the Internet. Machine learning models thus often adopt these biases when being trained on the bias-embedded datasets.
Technical flaws during data collection, sampling, and model design can further exacerbate unfairness by introducing edge cases, sampling bias, and temporal bias.
Therefore, the measurement of fairness with proper TAI metrics is critical for both SER and RecSys to guarantee that the users receive neutral and impartial services.

A common strategy of metric selection is to focus on one factor and measure the deviation from the equality of that factor. For example, SER normally focuses on the (potential) attention items receive from users in terms of factors such as click-through rates, exposure, or inferences of the content relevance. The deviation from equality for these factors can then be quantified via disparate impact, disparate exposure, disparate treatment~\cite{singh2018fairness,castillo2019fairness,zehlike2020reducing}, and inequity of attention~\cite{biega2018equity}. 
RecSys has also employed similar metrics for fairness, such as bias disparity, average disparity, and score disparity~\cite{tsintzou2018bias,leonhardt2018user}. Kullback–Leibler~(KL)-divergence has also been employed with ad-hoc adaptions to measure the fairness in SER~\cite{geyik2019fairness}.

Other issues also affect the definition and selection of metrics. For instance, fairness can refer to disparate treatment of individuals and of demographic groups, commonly termed as individual fairness~\cite{rastegarpanah2019fighting} and group fairness~\cite{kamishima2012enhancement}. The former can be seen as a special case of the latter where the group size equals one. 
We further discuss more similar matters in Section~\ref{sec:discussion}.

\paragraph{Privacy}
Service providers are consistently improving the user experience by providing personalized service using machine learning models trained with data about users' personal behavior and interests. A common method to acquire such data is to request permission to collect data when associating the search engine or RecSys services with a user account, e.g., Google and Youtube can be associated with the user's Google account. 
While improving user experience, this also raises privacy issues as personal information is sent to a remote server~\cite{xu2007privacy}.
Considering the volume and detail of data current systems collect, privacy concerns should be taken seriously~\cite{jeckmans2013privacy}. Additionally, companies often have financial incentives that conflict with protecting user privacy. For example, there has been a continuous series of privacy concerns over Google services. In 2012, Google changed its privacy policy to enable sharing data across a wide variety of services~\footnote{\url{https://policies.google.com/privacy/archive/20120301}}. While in 2016, Google quietly dropped its ban on personally-identifiable information in its DoubleClick ad service~\footnote{https://thetechportal.com/2016/10/21/google-now-personal-web-tracking-ads/}.

A common measurement logic is to assess the unwilling exposure of private information. For unstructured data like browsing history and email, entropy can provide a measure of unique information and quantify the amount of exposed private information~\cite{agrawal2001design}. Though, in practice, privacy preservation techniques can affect model performance. Therefore, privacy is commonly assessed as a multi-objective optimization problem. For example, cryptographic protocols, differential privacy, and anonymization approaches assess privacy preservation via trade-off measurements between privacy exposure (controlled by some characteristics of protocols or algorithms) and model performance using \emph{regression model metrics} and \emph{ranking metrics}~\cite{xu2007privacy,mcsherry2009differentially,xin2014controlling}.

\paragraph{Robustness}
As mentioned, enormous volumes of data are continuously generated online, thus models often require quite significant time and resources for retraining or incremental learning and therefore cannot always be done timely. This slow retraining and learning can result in issues after sudden data shifts. In addition to real data shifts, the problem of spamdexing, namely when users enter fake ratings to manipulate the ranking results, still exists in SER and related attacks are also common in collaborative filtering RecSys nowadays, dubbed as shilling attacks or profile injection attacks~\cite{lam2004shilling,williams2007defending}. Finally, malicious users may intentionally apply small targeted perturbations to datasets which can severely impact model performance, e.g., image classification tasks in multimedia recommender systems~\cite{moosavi2017universal}. Overall, an SER or RecSys is considered robust if not significantly affected by attacks like spamdexing and perturbations or dramatically data shifts~\cite{aggarwal2016recommender}. 

A common measurement method is to assess the model performance in the face of varying attacks or data shifts using \emph{ranking metrics}~\cite{wang2013theoretical,li2009incorporating,bailey2017retrieval,tang2019adversarial} and \emph{regression model metrics}~\cite{o2004collaborative,lam2004shilling,mobasher2006model,mobasher2007toward,cheng2010robust,gunes2014shilling}.

\subsection{Networking}
This section focuses on the networking domain and illustrates several example networking problems where different TAI metrics are in use. These examples are relevant for both wireless and wired networks and cover several different network layers. In contrast to the computing domain, most networking research considers only a single TAI metric area (as others are often considered irrelevant to the problem or out of scope); thus, each example problem also describes only a single TAI metric area. Relatedly, the metrics are more heterogeneous and ad-hoc than in the computing domain because TAI is very new in the networking domain; therefore, more detail and context are provided for each metric.

\paragraph{Wireless networking fairness.}
Radio resource allocation in wireless networks is a well-known networking problem with significant AI research and a major fairness component. Specifically, in the LTE context, this problem manifests as the allocation of physical resource blocks to specific user equipment on the sector level. Fairness in this context means ensuring that certain users are not starved of resources. 
Overall, such a problem is a multi-objective optimization problem with fairness as one objective with others such as high overall system performance (throughput). As network complexity and application diversity have risen, simple analytic or heuristic scheduling solutions (e.g., round-robin, proportional fair, and best CQI) are seen as potentially insufficient, and research has turned to reinforcement learning~(RL) to solve the problem. 

The fairness aspect in RL is thus embedded in the reward function as this function directs the learning. In the simplest cases, this fairness component of the reward function can be a weighted version of a traditional network fairness metric, such as Jain's fairness index or entropy, or a custom fairness metric resulting from reward engineering for the specific problem.
The basic Jain's fairness index is defined for a single type of network quality of service (QoS) measure (typically user throughput). Thus, the index does not consider further QoS measures (such as delay or packet loss) or users with different applications (and thus different QoS requirements). This index can be analogized to the independence group of algorithmic fairness measures as the index is blind (to such characteristics) and thus independent.

In more complex cases, such as with multiple QoS measures, different approaches are possible. For example Comsa et al. \shortcite{com20195mart} considers fairness (within a group of users of the same app) through a measure of the total sum of user-QoS requirement combinations (e.g., user $A$ with throughput threshold of $X$) met in a given period (one TTI). The authors then further generalizes this by considering many user groups with each group using another application. Specifically, the total fairness is a weighted sum of the intra-app-group fairnesses, with the weighting being a learnable (through RL) prioritization of the applications.
As another example, Al-Tam et al. \shortcite{al2020learn} use a variant of discounted best-CQI (a common network fairness model) as the reward function with the discounting based on the well known min-max ratio fairness measure.

\paragraph{Network traffic privacy.}
Network traffic classification is a major networking area, especially for mobile network operators as knowing the specific network traffic mixture supports many network tasks. For example, traffic mixture knowledge enables network optimizations including optimizing user QoS and QoE and better traffic volume prediction. However, more nefariously, traffic classification can also impinge on user privacy as entities (such as companies and governments) can use such classification to identify, for example, users of specific apps or websites (like those used by political dissidents). These network traffic classifiers often use a variety of ML and AI methods, including RF, SVM, and DNN, with and without handcrafted traffic features.
Thus, to counteract network traffic classification, researchers are applying both AI methods such as generative adversarial networks (GANs) \cite{li2019dynamic,fathi2020gan,hou2020wf} and non-AI methods, such as adaptive packet padding \cite{pinheiro2020adaptive} and optimized dummy packet injection \cite{shan2021patch}, to intelligently obfuscate the network traffic. These research studies primarily use related accuracy-based metrics to assess privacy improvement. 

Some works use the misclassification rate of the classifiers on the obfuscated traffic \cite{shan2021patch,hou2020wf}. Relatedly, other works use the differences in the classification accuracies of the classifier on the original and obfuscated traffic \cite{pinheiro2020adaptive}. Goal-wise, certain studies analyze the classifier accuracy, recall, and precision on traffic of one type disguised (by a GAN) to look like traffic of a different (target) type (rather than a goal of just general obfuscation) \cite{fathi2020gan,hou2020wf}. Finally, Li et al. \shortcite{li2019dynamic} uses both indistinguishability under Classification Attack (IND-CA) and the differences in AUC of the ROC curve of the classifier on the original and obfuscated traffic. IND-CA quantifies how distinguishable the traffic is when considering two traffic types with equal shares (each representing 50\% of traffic). Specifically, IND-CA represents the normalized benefit of the classification with zero meaning a random guess and, in contrast, one meaning full certainty.

\paragraph{Networking congestion control robustness.}
Similar to the resource allocation problem, the networking congestion control problem also lends well to an RL approach. The issue also deals with significant multi-layer network complexity (where simpler heuristics are insufficient). Specifically, the major target is TCP congestion control with RL adapting the receive window size. 
A potential benefit to the RL approach is its robustness to different network conditions. 

However, this is sometimes only the case if those network conditions were part of the training regime (for offline RL). Thus, robustness testing with conditions that both span and extend beyond the training regime is important. Some of the TCP RL works use regret-based metrics with a baseline scenario containing only network conditions from the training range. The metric is then defined as the performance gap between this baseline and scenarios that include conditions beyond the training regime \cite{he2021deepcc}.

In other cases where robustness to different conditions is built-in to the RL approach, the metric is simply the performance gap in the diverse conditions of this approach from baseline approaches. Du et al. \shortcite{du2021unified}, for example, use a hybrid approach with traditional (heuristic) and RL parts to improve robustness in both diverse wired and wireless network situations. The approach proves better (in terms of mean throughput and delay) than either approach alone in such situations.
Other works use statistical dispersion metrics such as standard deviation and confidence intervals to illustrate the general robustness of the results (for example, where stability is important). For instance, an RL approach \cite{xiao2019tcp} illustrates a 95\% confidence interval that is smaller than all the baselines for a specific performance measure (a fairness index); thus illustrating robustness to more significant swings.

\section{Lessons Learned and Research Agenda}\label{sec:discussion}

By examining the TAI metric selection across computing and networking domains, we can see that the definition and selection of TAI metrics for computing are more straightforward than for networking. A significant reason is that the outputs of many computing systems like recommendation systems and search engines are more oriented towards end-users, e.g., the results of ranking algorithms well match the needs of SER which uses user data and show results directly to users. 

In contrast, the learning algorithms in the networking context usually result in intermediate metrics that serve to adapt protocols or algorithms that eventually affect the target metrics. In other words, the output of the learning algorithms in computing can often be directly used to assess user experience, while in networking there is normally an intermediate model to transfer networking performance (controlled by AI) to user experience. Thus, TAI metrics in networking require more ad-hoc designs, definitions of usage and user context, and targets of networking systems.

Currently, most computing and networking systems use ``functionality-driven design'', which we use as a contrast to ``user-centered design'', in the sense that the former focuses more on pre-defined systematic performance metrics though also sometimes considers user-related metrics, such as QoS and QoE. Relatedly, TAI design and metric selection in such systems also often focus on the most functionalities during specific life-cycle phases. Ideally, TAI metric selection demands more thorough considerations to guarantee trustworthiness through the life cycle of usage.
Furthermore, the current mindset of TAI design and metric selection, restricted by the aforementioned design philosophies, takes into consideration only part of human cognition, specifically the conscious and concrete areas that can be more easily measured and quantified, such as pattern recognition, language, attention, perception, and action. These are widely explored by AI communities. 

However, the exploration of the unconscious and abstract areas of cognition, e.g., mental health and emotion, is just beginning. Methodological limits is a key reason for this, e.g., lack of devices and theories to accurately capture bioelectrical signals and convert these signals to emotional statuses.

Trustworthiness itself consists of cognitive, emotional and behavioral factors, since trustworthiness is a user-oriented term. This aspect will play an increasingly important role when ``user-centered design'' dominates future cyberspace, replacing the current ``functionality-driven design''. In the future, considering the unconscious and abstract cognition areas will be vital to guarantee TAI. These parts are hard to quantify and might remain so even with advanced techniques in sensor-enabled immersive cyberspace. Therefore, other assessment methods may be required for TAI in such immersive cyberspaces. This is discussed in the remaining paragraphs. 

\paragraph{From ad-hoc to systematic metric selection}
Although we presented an outlook on how TAI metrics are selected from the surveyed literature in computing and networking domains, it can be more complicated when system developers try to select or define TAI metrics in reality. Because achieving trustworthiness highly relies on the contextual environment besides the AI system itself, the selection of metrics requires a holistic and systemic consideration encompassing all processes within the system’s socio-technical context throughout its entire life cycle. 

Therefore, every system, even if it is an identical copy of another, may require different TAI metrics or some metric adaptions due to the differences in the deployed contextual environments and life cycles. Moreover, the selection of metrics also depends on the granularity of the concerned target (pointwise, pairwise, listwise) and the operation phase (pre-processing, in-processing, post-processing). Thus, even for straightforward computing tasks, developers need delicate considerations about the whole context and potential scenarios to fully justify the selection of TAI metrics.

\paragraph{TAI metric selection for immersive cyberspace} 

Current cyberspace is evolving as new technologies develop, and the advent of immersive cyberspace will be enabled by AI in a greater extent. For example, since 2021, the metaverse, also known as \textit{the immersive Internet}, has risen to public attention, thanks to Facebook's rebranding, and worldwide academic and industrial promotion. In the metaverse, AI continues to play a core role as the foundation of several key technologies, namely, computer vision, augmented \& virtual reality (AR/VR), data mining, and robotics~\cite{lee2021all}. The most critical difference between the metaverse and current cyberspace is that in the metaverse, human users are absorbed into the projected blended virtual-physical world without explicit exit points instead of just standing by as external interactors. As such, any consequences caused by AI misbehavior could be significantly worse and hence severely impact human users' welfare. 

For example, VR technologies are capable of recording richer personal data, such as eye movements and emotional reactions, which could be deployed in threatening ways powered by AI techniques to manipulate users’ beliefs, emotions, and behaviors~\cite{spiegel2018ethics}. In contrast, AR requires strong contextual awareness, in terms of users, their adjacent environments, and social interactions, to augment the physical world~\cite{10.1145/3474085.3475413}. As such, users have to share egocentric views of various contexts, e.g., Project EGO4D\footnote{\url{https://ego4d-data.org/}}. 

In other words, users have to build trust with virtual-physical blended cyberspace mediated by computing systems, as the users' daily interactions with physical worlds and people are recorded in an unprecedentedly massive scale. Meanwhile, users will require an easy-to-interpret score, e.g., AI Trust Score~\cite{10.1145/3411763.3443452}, to judge AI trustworthiness over time (instead of a single snapshot), while AI is resilient to the occurrence of glitches and afterwards able to recover trusts with users.
Furthermore, the ``user-centered'' features of the metaverse may bring important changes to TAI and TAI metric selection. In current cyberspace, AI-enabled applications and human users interact but are still significantly and explicitly separated. Hence the measure of TAI mainly focuses on system performance and technical metrics. 

In the metaverse, however, user-representative avatars, cognitive emotional-interactive products, and other similar humanoids will play vital roles in improving a user's feeling of involvement to seamlessly experience the blended virtual-physical world. More importantly, the aforementioned avatars and humanoids will collaborate with human users. Thus, user-centered TAI metrics, such as those focusing on cognition, sentiment, and psychology, with sensor-enabled monitoring in the metaverse, will become a new driver for understanding robustness, privacy and fairness.

Emerging techniques may have the potential to tackle the challenge through understanding the internal states of users. For example, the state-of-the-art Brain-Computer Interface (BCI) can estimate the user's current emotion, attention, or fatigue level to some extent by monitoring the bioelectrical signals that reflect brain activity. These signals can be recorded by a device like an electroencephalogram~\cite{10.1145/3457950}. These emerging techniques may allow for quantitatively measuring the abstract metrics that currently rely on limited-scale qualitative experiments (often based on user interviews).

Nevertheless, these techniques are still immature. Moreover, the practicality and applicability of the techniques maybe be limited, as they normally require the users to wear additional devices, which are often inconvenient. Nowadays, immersive headsets (AR/VR) have similar issues~\cite{Lee2022TowardsAR}. Therefore, requirements for qualitative measures like user studies for TAI assessment, e.g., eliciting user requirements for trust-guarantee AI services, might be reasonable. However, the current approaches for understanding users, to a large extent, are costly and time-consuming. Thus, this timeliness issue is a current challenge to be solved. 

\paragraph{TAI governance}
Currently, the governance of the trustworthiness of AI systems is mostly left in the hands of the service providers themselves and some third-party companies and government institutes. As user awareness of TAI increases, a key challenge for TAI governance is how to assess and guarantee fairness, privacy, and robustness in a more standardized, transparent, and systematic way. Moreover, the coming metaverse will integrate more AI services into daily life on an unprecedentedly massive scale. Governance by each individual service provider creates coordination and standardization problems and thus provides no guarantee of equity in trustworthiness standards. 
Additionally, the burden might be too large for a limited number of third party companies or government institutes.
Accordingly, building an autonomous governance
platform, ``meta-TAI'', to govern TAI performance, might be worth exploring. The meta-TAI platform could be collaboratively governed by a number of trustworthy institutes authorized by the involved countries, as well as offering TAI scores for various providers of AI services and their individual solutions. As such, the platform can save manual effort while ensuring the equity of standards across various phases of a user-AI interaction cycle.

\section{Conclusion}
This survey discussed metrics for TAI and further examined the aspects of fairness, privacy and robustness in the computing and networking domains. The existing metrics are mainly driven by system functionalities and efficacy with less emphasis on user-centered factors. Meanwhile, the ad-hoc metric selection causes sub-optimal results in building trustworthiness with users. 
We revisited the TAI domain to lay out a research agenda that will assist researchers working on TAI and immersive cyberspace to contextualize and focus their efforts. We note that AI will become an indispensable driver of immersive cyberspace and that users will interact with AI-enabled services in this virtual-physical blended world under the premise that user trust is essential to the wide adoption of such services. Therefore, we call for a user-centered paradigm of building trustworthiness beyond sole system measurements and considering cognitive and affective factors.

\bibliographystyle{named}
\bibliography{trim}

\begin{thebibliography}{}

\bibitem[\protect\citeauthoryear{Adadi and Berrada}{2018}]{adadi2018peeking}
A.~Adadi and M.~Berrada.
\newblock Peeking inside the black-box: A survey on explainable artificial
  intelligence (xai).
\newblock {\em IEEE Access}, 6:52138--52160, 2018.

\bibitem[\protect\citeauthoryear{Aggarwal}{2016}]{aggarwal2016recommender}
C.~C. Aggarwal.
\newblock {\em Recommender systems}, volume~1.
\newblock Springer, 2016.

\bibitem[\protect\citeauthoryear{Agrawal and
  Aggarwal}{2001}]{agrawal2001design}
D.~Agrawal and C.~C. Aggarwal.
\newblock On the design and quantification of privacy preserving data mining
  algorithms.
\newblock In {\em Proc. of the 20th ACM SIGMOD-SIGACT-SIGART Symp. on
  Principles of database systems}, pages 247--255, 2001.

\bibitem[\protect\citeauthoryear{Al-Tam \bgroup \em et al.\egroup
  }{2020}]{al2020learn}
F.~Al-Tam, N.~Correia, and J.~Rodriguez.
\newblock Learn to schedule (leasch): A deep reinforcement learning approach
  for radio resource scheduling in the 5g mac layer.
\newblock {\em IEEE Access}, 8:108088--108101, 2020.

\bibitem[\protect\citeauthoryear{Bailey \bgroup \em et al.\egroup
  }{2017}]{bailey2017retrieval}
P.~Bailey, A.~Moffat, F.~Scholer, and P.~Thomas.
\newblock Retrieval consistency in the presence of query variations.
\newblock In {\em Proc. of the 40th Inter. ACM SIGIR}, pages 395--404, 2017.

\bibitem[\protect\citeauthoryear{Ben-Tal \bgroup \em et al.\egroup
  }{2009}]{ben2009robust}
A.~Ben-Tal, L.~El~Ghaoui, and A.~Nemirovski.
\newblock {\em Robust optimization}.
\newblock Princeton Uni. press, 2009.

\bibitem[\protect\citeauthoryear{Biega \bgroup \em et al.\egroup
  }{2018}]{biega2018equity}
A.~J. Biega, K.~P. Gummadi, and G.~Weikum.
\newblock Equity of attention: Amortizing individual fairness in rankings.
\newblock In {\em The 41st Inter. ACM SIGIR Conf.}, pages 405--414, 2018.

\bibitem[\protect\citeauthoryear{Castillo}{2019}]{castillo2019fairness}
C.~Castillo.
\newblock Fairness and transparency in ranking.
\newblock In {\em ACM SIGIR Forum}, volume~52, pages 64--71. ACM USA, 2019.

\bibitem[\protect\citeauthoryear{Cheng and Hurley}{2010}]{cheng2010robust}
Z.~Cheng and N.~Hurley.
\newblock Robust collaborative recommendation by least trimmed squares matrix
  factorization.
\newblock In {\em 2010 22nd IEEE Inter. Conf. on Tools with AI}, volume~2,
  pages 105--112. IEEE, 2010.

\bibitem[\protect\citeauthoryear{Cheng \bgroup \em et al.\egroup
  }{2016}]{cheng2016wide}
H.-T. Cheng, L.~Koc, J.~Harmsen, et~al.
\newblock Wide \& deep learning for recommender systems.
\newblock In {\em Proc. of the 1st WKSP on deep learning for recommender
  systems}, pages 7--10, 2016.

\bibitem[\protect\citeauthoryear{Comșa \bgroup \em et al.\egroup
  }{2019}]{com20195mart}
I.-S. Comșa, R.~Trestian, G.-M. Muntean, and G.~Ghinea.
\newblock 5mart: A 5g smart scheduling framework for optimizing qos through
  reinforcement learning.
\newblock {\em IEEE Trans. on Network and Service Mgt.}, 17(2):1110--1124,
  2019.

\bibitem[\protect\citeauthoryear{Cui \bgroup \em et al.\egroup
  }{2006}]{cui2006machine}
G.~Cui, M.-L. Wong, and H.-K. Lui.
\newblock Machine learning for direct marketing response models: Bayesian
  networks with evolutionary programming.
\newblock {\em Mgt. Sci.}, 52(4):597--612, 2006.

\bibitem[\protect\citeauthoryear{Du \bgroup \em et al.\egroup
  }{2021}]{du2021unified}
Z.~Du, J.~Zheng, H.~Yu, L.~Kong, and G.~Chen.
\newblock A unified congestion control framework for diverse application
  preferences and network conditions.
\newblock In {\em Proc. of the 17th Inter. Conf. on emerging Networking
  Experiments and Tech.}, pages 282--296, 2021.

\bibitem[\protect\citeauthoryear{{European Union}}{2012}]{euhumanright}
{European Union}.
\newblock Eu charter of fundamental rights.
\newblock 2012.

\bibitem[\protect\citeauthoryear{Fathi-Kazerooni and
  Rojas-Cessa}{2020}]{fathi2020gan}
S.~Fathi-Kazerooni and R.~Rojas-Cessa.
\newblock Gan tunnel: network traffic steganography by using gans to counter
  internet traffic classifiers.
\newblock {\em IEEE Access}, 8:125345--125359, 2020.

\bibitem[\protect\citeauthoryear{Geyik \bgroup \em et al.\egroup
  }{2019}]{geyik2019fairness}
S.~C. Geyik, S.~Ambler, and K.~Kenthapadi.
\newblock Fairness-aware ranking in search \& recommendation systems with
  application to linkedin talent search.
\newblock In {\em Proc. of the 25th ACM SIGKDD}, pages 2221--2231, 2019.

\bibitem[\protect\citeauthoryear{Gunes \bgroup \em et al.\egroup
  }{2014}]{gunes2014shilling}
I.~Gunes, C.~Kaleli, A.~Bilge, and H.~Polat.
\newblock Shilling attacks against recommender systems: a comprehensive survey.
\newblock {\em AI Review}, 42(4):767--799, 2014.

\bibitem[\protect\citeauthoryear{He \bgroup \em et al.\egroup
  }{2021}]{he2021deepcc}
B.~He, J.~Wang, Q.~Qi, et~al.
\newblock Deepcc: Multi-agent deep reinforcement learning congestion control
  for multi-path tcp based on self-attention.
\newblock {\em IEEE Trans. on Network and Service Mgt.}, 18(4):4770--4788,
  2021.

\bibitem[\protect\citeauthoryear{Heaven}{2021}]{usdod}
W.~D. Heaven.
\newblock The department of defense is issuing ai ethics guidelines for tech
  contractors.
\newblock 2021.

\bibitem[\protect\citeauthoryear{{HLEGAI}}{2019}]{hleg2019ethics}
{HLEGAI}.
\newblock Ethics guidelines for trustworthy ai.
\newblock {\em B-1049 Brussels}, 2019.

\bibitem[\protect\citeauthoryear{Hou \bgroup \em et al.\egroup
  }{2020}]{hou2020wf}
C.~Hou, G.~Gou, et~al.
\newblock Wf-gan: Fighting back against website fingerprinting attack using
  adversarial learning.
\newblock In {\em 2020 IEEE Symp.on Comp. and Commun. (ISCC)}, pages 1--7.
  IEEE, 2020.

\bibitem[\protect\citeauthoryear{Jeckmans \bgroup \em et al.\egroup
  }{2013}]{jeckmans2013privacy}
A.~JP Jeckmans, M.~Beye, Z.~Erkin, et~al.
\newblock Privacy in recommender systems.
\newblock In {\em Social media retrieval}, pages 263--281. Springer, 2013.

\bibitem[\protect\citeauthoryear{Joachims \bgroup \em et al.\egroup
  }{2017}]{joachims2017accurately}
T.~Joachims, L.~Granka, B.~Pan, H.~Hembrooke, and G.~Gay.
\newblock Accurately interpreting clickthrough data as implicit feedback.
\newblock In {\em ACM SIGIR Forum}, volume~51, pages 4--11. ACM USA, 2017.

\bibitem[\protect\citeauthoryear{Jordan and Mitchell}{2015}]{jordan2015machine}
M.~I. Jordan and T.~M Mitchell.
\newblock Machine learning: Trends, perspectives, and prospects.
\newblock {\em Science}, 349(6245):255--260, 2015.

\bibitem[\protect\citeauthoryear{Kamishima \bgroup \em et al.\egroup
  }{2012}]{kamishima2012enhancement}
T.~Kamishima, S.~Akaho, H.~Asoh, et~al.
\newblock Enhancement of the neutrality in recommendation.
\newblock In {\em Decisions@ RecSys}, pages 8--14. Citeseer, 2012.

\bibitem[\protect\citeauthoryear{Lam and Riedl}{2004}]{lam2004shilling}
S.~K. Lam and J.~Riedl.
\newblock Shilling recommender systems for fun and profit.
\newblock In {\em Proc. of the 13th Inter. Conf. on WWW}, pages 393--402, 2004.

\bibitem[\protect\citeauthoryear{Lam \bgroup \em et al.\egroup
  }{2021}]{10.1145/3474085.3475413}
K.~Y. Lam, L.~H. Lee, and P.~Hui.
\newblock {\em A2W: Context-Aware Recommendation System for Mobile Augmented
  Reality Web Browser}, page 2447–2455.
\newblock ACM, USA, 2021.

\bibitem[\protect\citeauthoryear{Laufer}{1973}]{laufer1973some}
R.~S. Laufer.
\newblock Some analytic dimensions of priacy.
\newblock In {\em Architectural psychology, Proc. of the Lund Conf., Lund:
  Studentlitteratur}, 1973.

\bibitem[\protect\citeauthoryear{Lee \bgroup \em et al.\egroup
  }{2021}]{lee2021all}
L.-H. Lee, T.~Braud, P.~Zhou, et~al.
\newblock All one needs to know about metaverse: A complete survey on
  technological singularity, virtual ecosystem, and research agenda.
\newblock {\em arXiv:2110.05352}, 2021.

\bibitem[\protect\citeauthoryear{Lee \bgroup \em et al.\egroup
  }{2022}]{Lee2022TowardsAR}
L.-H. Lee, T.~Braud, S.~Hosio, and P.~Hui.
\newblock Towards augmented reality driven human-city interaction: Current
  research on mobile headsets and future challenges.
\newblock {\em ACM CSUR}, 54:1 -- 38, 2022.

\bibitem[\protect\citeauthoryear{Leonhardt \bgroup \em et al.\egroup
  }{2018}]{leonhardt2018user}
J.~Leonhardt, A.~Anand, and M.~Khosla.
\newblock User fairness in recommender systems.
\newblock In {\em Compan. Proc. of the The Web Conf. 2018}, pages 101--102,
  2018.

\bibitem[\protect\citeauthoryear{Li \bgroup \em et al.\egroup
  }{2009}]{li2009incorporating}
X.~Li, F.~Li, S.~Ji, Z.~Zheng, et~al.
\newblock Incorporating robustness into web ranking evaluation.
\newblock In {\em Proc. of the 18th ACM Conf. on Info. and Knowledge Mgt.},
  pages 2007--2010, 2009.

\bibitem[\protect\citeauthoryear{Li \bgroup \em et al.\egroup
  }{2019}]{li2019dynamic}
J.~Li, L.~Zhou, H.~Li, L.~Yan, and H.~Zhu.
\newblock Dynamic traffic feature camouflaging via generative adversarial
  networks.
\newblock In {\em 2019 IEEE Conf. on Comm. and Network Security (CNS)}, pages
  268--276. IEEE, 2019.

\bibitem[\protect\citeauthoryear{Lin \bgroup \em et al.\egroup
  }{2011}]{lin2011machine}
W.-Y. Lin, Y.-H. Hu, and C.-F. Tsai.
\newblock Machine learning in financial crisis prediction: a survey.
\newblock {\em IEEE Trans. on Systems, Man, and Cybernetics, Part C
  (Applications and Reviews)}, 42(4):421--436, 2011.

\bibitem[\protect\citeauthoryear{Maier \bgroup \em et al.\egroup
  }{2016}]{maier2016uncertain}
H.~R. Maier, J.~HA Guillaume, H.~van Delden, et~al.
\newblock An uncertain future, deep uncertainty, scenarios, robustness and
  adaptation: How do they fit together?
\newblock {\em Environ. Modelling \& Soft.}, 81:154--164, 2016.

\bibitem[\protect\citeauthoryear{Mantelero}{2018}]{mantelero2018ai}
A.~Mantelero.
\newblock Ai and big data: A blueprint for a human rights, social and ethical
  impact assessment.
\newblock {\em Comp. Law \& Security Review}, 34(4):754--772, 2018.

\bibitem[\protect\citeauthoryear{McSherry and
  Mironov}{2009}]{mcsherry2009differentially}
F.~McSherry and I.~Mironov.
\newblock Differentially private recommender systems: Building privacy into the
  netflix prize contenders.
\newblock In {\em Proc. of the 15th ACM SIGKDD}, pages 627--636, 2009.

\bibitem[\protect\citeauthoryear{Mobasher \bgroup \em et al.\egroup
  }{2006}]{mobasher2006model}
B.~Mobasher, R.~Burke, and J.~J. Sandvig.
\newblock Model-based collaborative filtering as a defense against profile
  injection attacks.
\newblock In {\em AAAI}, volume~6, page 1388, 2006.

\bibitem[\protect\citeauthoryear{Mobasher \bgroup \em et al.\egroup
  }{2007}]{mobasher2007toward}
B.~Mobasher, R.~Burke, R.~Bhaumik, and C.~Williams.
\newblock Toward trustworthy recommender systems: An analysis of attack models
  and algorithm robustness.
\newblock {\em ACM TOIT}, 7(4):23--es, 2007.

\bibitem[\protect\citeauthoryear{Moosavi-Dezfooli \bgroup \em et al.\egroup
  }{2017}]{moosavi2017universal}
S.-M. Moosavi-Dezfooli, A.~Fawzi, O.~Fawzi, and P.~Frossard.
\newblock Universal adversarial perturbations.
\newblock In {\em Proc. of the IEEE Conf. on CVPR}, pages 1765--1773, 2017.

\bibitem[\protect\citeauthoryear{Nissenbaum}{2004}]{nissenbaum2004privacy}
H.~Nissenbaum.
\newblock Privacy as contextual integrity.
\newblock {\em Wash. L. Rev.}, 79:119, 2004.

\bibitem[\protect\citeauthoryear{Norgeot \bgroup \em et al.\egroup
  }{2019}]{norgeot2019call}
B.~Norgeot, B.~S. Glicksberg, and A.~J. Butte.
\newblock A call for deep-learning healthcare.
\newblock {\em Nature medicine}, 25(1):14--15, 2019.

\bibitem[\protect\citeauthoryear{O'Mahony \bgroup \em et al.\egroup
  }{2004}]{o2004collaborative}
M.~O'Mahony, N.~Hurley, N.~Kushmerick, and G.~Silvestre.
\newblock Collaborative recommendation: A robustness analysis.
\newblock {\em ACM TOIT}, 4(4):344--377, 2004.

\bibitem[\protect\citeauthoryear{Paluck \bgroup \em et al.\egroup
  }{2019}]{paluck2019contact}
E.~L. Paluck, S.~A. Green, and D.~P. Green.
\newblock The contact hypothesis re-evaluated.
\newblock {\em Behavioural Public Policy}, 3(2):129--158, 2019.

\bibitem[\protect\citeauthoryear{Pinheiro \bgroup \em et al.\egroup
  }{2020}]{pinheiro2020adaptive}
A.~J. Pinheiro, P.~F. de~Araujo-Filho, J.~de~M. Bezerra, and D.~R. Campelo.
\newblock Adaptive packet padding approach for smart home networks: A tradeoff
  between privacy and performance.
\newblock {\em IEEE Internet of Things J.}, 8(5):3930--3938, 2020.

\bibitem[\protect\citeauthoryear{Rastegarpanah \bgroup \em et al.\egroup
  }{2019}]{rastegarpanah2019fighting}
B.~Rastegarpanah, K.~P. Gummadi, and M.~Crovella.
\newblock Fighting fire with fire: Using antidote data to improve polarization
  and fairness of recommender systems.
\newblock In {\em Proc. of the 12th ACM Inter. Conf. on WSDM}, pages 231--239,
  2019.

\bibitem[\protect\citeauthoryear{Shan \bgroup \em et al.\egroup
  }{2021}]{shan2021patch}
S.~Shan, A.~N. Bhagoji, H.~Zheng, and B.~Y. Zhao.
\newblock Patch-based defenses against web fingerprinting attacks.
\newblock In {\em Proc. of the 14th ACM WKSP on AI and Security}, pages
  97--109, 2021.

\bibitem[\protect\citeauthoryear{Shatilov \bgroup \em et al.\egroup
  }{2021}]{10.1145/3457950}
K.~A. Shatilov, D.~Chatzopoulos, L.-H. Lee, and P.~Hui.
\newblock Emerging exg-based nui inputs in extended realities: A bottom-up
  survey.
\newblock {\em ACM Trans. Interact. Intell. Syst.}, 11(2), jul 2021.

\bibitem[\protect\citeauthoryear{Shiue \bgroup \em et al.\egroup
  }{2018}]{shiue2018real}
Y.-R. Shiue, K.-C. Lee, and C.-T. Su.
\newblock Real-time scheduling for a smart factory using a reinforcement
  learning approach.
\newblock {\em Computers \& Industrial Eng.}, 125:604--614, 2018.

\bibitem[\protect\citeauthoryear{Singh and Joachims}{2018}]{singh2018fairness}
A.~Singh and T.~Joachims.
\newblock Fairness of exposure in rankings.
\newblock In {\em Proc. of the 24th ACM SIGKDD}, pages 2219--2228, 2018.

\bibitem[\protect\citeauthoryear{Smuha}{2019}]{smuha2019eu}
N.~A. Smuha.
\newblock The eu approach to ethics guidelines for trustworthy artificial
  intelligence.
\newblock {\em CRi-Comp. Law Review Inter.}, 2019.

\bibitem[\protect\citeauthoryear{Spiegel}{2018}]{spiegel2018ethics}
J.~S. Spiegel.
\newblock The ethics of virtual reality technology: Social hazards and public
  policy recommendations.
\newblock {\em Sci. and Eng. ethics}, 24(5):1537--1550, 2018.

\bibitem[\protect\citeauthoryear{Tang \bgroup \em et al.\egroup
  }{2019}]{tang2019adversarial}
J.~Tang, X.~Du, X.~He, F.~Yuan, et~al.
\newblock Adversarial training towards robust multimedia recommender system.
\newblock {\em IEEE Trans. on KDE}, 32(5):855--867, 2019.

\bibitem[\protect\citeauthoryear{Tavani}{2012}]{tavani2012search}
H.~Tavani.
\newblock Search engines and ethics.
\newblock 2012.

\bibitem[\protect\citeauthoryear{Torresen}{2018}]{torresen2018review}
J.~Torresen.
\newblock A review of future and ethical perspectives of robotics and ai.
\newblock {\em Frontiers in Robotics and AI}, 4:75, 2018.

\bibitem[\protect\citeauthoryear{Tsintzou \bgroup \em et al.\egroup
  }{2018}]{tsintzou2018bias}
V.~Tsintzou, E.~Pitoura, and P.~Tsaparas.
\newblock Bias disparity in recommendation systems.
\newblock {\em arXiv:1811.01461}, 2018.

\bibitem[\protect\citeauthoryear{Wang and
  Moulden}{2021}]{10.1145/3411763.3443452}
J.~Wang and A.~Moulden.
\newblock {\em AI Trust Score: A User-Centered Approach to Building, Designing,
  and Measuring the Success of Intelligent Workplace Features}.
\newblock ACM, USA, 2021.

\bibitem[\protect\citeauthoryear{Wang \bgroup \em et al.\egroup
  }{2013}]{wang2013theoretical}
Y.~Wang, L.~Wang, Y.~Li, et~al.
\newblock A theoretical analysis of ndcg ranking measures.
\newblock In {\em Proc. of the 26th Conf. COLT}, volume~8, page~6. Citeseer,
  2013.

\bibitem[\protect\citeauthoryear{Williams \bgroup \em et al.\egroup
  }{2007}]{williams2007defending}
C.~A. Williams, B.~Mobasher, and R.~Burke.
\newblock Defending recommender systems: detection of profile injection
  attacks.
\newblock {\em Service Oriented Comp. \& Apps.}, 1(3):157--170, 2007.

\bibitem[\protect\citeauthoryear{Xiao \bgroup \em et al.\egroup
  }{2019}]{xiao2019tcp}
K.~Xiao, S.~Mao, and J.~K. Tugnait.
\newblock Tcp-drinc: Smart congestion control based on deep reinforcement
  learning.
\newblock {\em IEEE Access}, 7:11892--11904, 2019.

\bibitem[\protect\citeauthoryear{Xin and Jaakkola}{2014}]{xin2014controlling}
Y.~Xin and T.~Jaakkola.
\newblock Controlling privacy in recommender systems.
\newblock NeurIPS, 2014.

\bibitem[\protect\citeauthoryear{Xu \bgroup \em et al.\egroup
  }{2007}]{xu2007privacy}
Y.~Xu, K.~Wang, B.~Zhang, and Z.~Chen.
\newblock Privacy-enhancing personalized web search.
\newblock In {\em Proc. of the 16th Inter. Conf. on WWW}, pages 591--600, 2007.

\bibitem[\protect\citeauthoryear{Zehlike and
  Castillo}{2020}]{zehlike2020reducing}
M.~Zehlike and C.~Castillo.
\newblock Reducing disparate exposure in ranking: A learning to rank approach.
\newblock In {\em Proc. of The Web Conf. 2020}, pages 2849--2855, 2020.

\end{thebibliography}

\end{document}